\documentclass[aps,reprint,twocolumn,superscriptaddress,longbibliography,prl]{revtex4-1}
\usepackage{amsmath,amssymb,amsfonts}
\usepackage{graphicx}
\usepackage{bm}
\usepackage{dsfont}
\usepackage[usenames,dvipsnames,svgnames,table]{xcolor}
\usepackage[colorlinks=True,linkcolor=red,citecolor=blue,urlcolor=blue]{hyperref}
\usepackage{hypcap}
\usepackage{ulem}

\newcommand{\ket}[1]{\left\vert #1 \right\rangle}
\newcommand{\bra}[1]{\left\langle #1 \right\vert}

\newcommand{\mrm}[1]{\mathrm{#1}}

\def\be{\begin{equation}}
\def\ee{\end{equation}}
\def\bea{\begin{eqnarray}}
\def\eea{\end{eqnarray}}

\newcommand{\diff}{\boldsymbol{\nabla}}

\def\ket#1{|#1\rangle }
\def\bra#1{\langle #1 |}

\begin{document}

\title{
Landau levels, Bardeen polynomials and Fermi arcs in Weyl semimetals:\\
the who's who of the chiral anomaly 
}
\author{Jan Behrends}
\affiliation{Max-Planck-Institut f{\"u}r Physik komplexer Systeme, 01187 Dresden, Germany}
\author{Sthitadhi Roy}
\affiliation{Physical and Theoretical Chemistry, Oxford University, South Parks Road, Oxford OX1 3QZ, United Kingdom}
\affiliation{Rudolf Peierls Centre For Theoretical Physics, Clarendon Laboratory, Oxford University, Parks Road, Oxford OX1 3PU, United Kingdom}
\affiliation{Max-Planck-Institut f{\"u}r Physik komplexer Systeme, 01187 Dresden, Germany}
\author{Michael H. Kolodrubetz}
\affiliation{Department of Physics, University of California, Berkeley, CA 94720, USA}
\affiliation{Materials Sciences Division, Lawrence Berkeley National Laboratory, Berkeley, CA 94720, USA}
\affiliation{Department of Physics, The University of Texas at Dallas, Richardson, TX
  75080, USA}
\author{Jens H. Bardarson}
\affiliation{Department of Physics, KTH Royal Institute of Technology, Stockholm, SE-106 91 Sweden}
\affiliation{Max-Planck-Institut f{\"u}r Physik komplexer Systeme, 01187 Dresden, Germany}
\author{Adolfo G. Grushin}
\affiliation{Department of Physics, University of California, Berkeley, CA 94720, USA}
\affiliation{Institut N{\'e}el, CNRS and Universit{\'e} Grenoble Alpes, F-38042 Grenoble, France}
\begin{abstract}
Condensed matter systems realizing Weyl fermions exhibit striking phenomenology derived from their topologically protected surface states as well as chiral anomalies induced by electromagnetic fields.
More recently, inhomogeneous strain or magnetization were predicted to result in chiral electric $\mathbf{E}_5$ and magnetic $\mathbf{B}_5$ fields, which modify and enrich the chiral anomaly with additional terms. 
In this work 
we develop a lattice-based approach to describe the chiral anomaly, which involves Landau and pseudo-Landau levels and treats all anomalous terms on equal footing, while naturally incorporating Fermi arcs.
We exemplify its potential by physically interpreting the largely overlooked role of Fermi arcs in the covariant (Fermi level) contribution to the anomaly
and revisiting the factor of $1/3$ difference between the covariant and consistent (complete band) contributions to the $\mathbf{E}_5\cdot\mathbf{B}_5$ term in the anomaly.
Our framework provides a versatile tool for the analysis of anomalies in realistic lattice models as well as a source of simple physical intuition for understanding strained and magnetized inhomogeneous Weyl semimetals.
\end{abstract}
\maketitle


Quantum anomalies describe the breaking of a classical symmetry by quantum fluctuations~\cite{bertlmann2000anomalies}.
The chiral anomaly, the nonconservation of the chiral charge of three-dimensional Weyl fermions, is relevant to different domains in physics since Weyl fermions mediate the pion-decay into photons~\cite{Adler:1969ir,Bell:1969dd} and are emergent quasiparticles in Weyl semimetals~\cite{Turner2013,Hosur2013,Miransky2015,Jia2016,Burkov2016,Armitage2017}.
The physics is particularly transparent in the Landau level picture pioneered by Nielsen and Ninomiya~\cite{NielNino83}, requiring only basic quantum mechanics.
In a magnetic field $\mathbf{B}$, the conical Weyl dispersion evolves into Landau levels with a degeneracy proportional to $|\mathbf{B}|$~\cite{NielNino83}.
Since momentum along the direction of the magnetic field remains a good quantum number, the Landau levels disperse in that direction, with the zeroth Landau level having a linear dispersion with a sign determined by the chirality;
all other Landau levels have a quadratic dispersion.
The zeroth Landau level of the left- and right-handed Weyl fermions furthermore connect at high energy.
Consequently, an electric field ($\mathbf{E}$) with a component along the dispersion transfers left-handed fermions to right-handed fermions (or vice versa) resulting in a nonconservation of left and right particle numbers proportional to $\mathbf{E}\cdot\mathbf{B}$~\cite{bertlmann2000anomalies,H93,NielNino83}.

Other fields, such as chiral pseudo-electromagnetic fields, torsion or curvature activate the chiral anomaly beyond $\mathbf{E}$ and $\mathbf{B}$~\cite{Parrikar:2014bv,Sun:2014ca,Zubkov:2015hk,Cortijo2015,PCF16,GVV16,You:2016hk,Landsteiner:2011hy,Gooth2017,Ferreiros:2019kz,Lepori:2018gb}.
Weyl semimetals are ideal to probe the chiral anomaly in the presence of chiral pseudo-electromagnetic fields.
To motivate this, recall that their low-energy degrees of freedom are pairs of chiral Weyl quasiparticles at topologically protected band touchings (Weyl nodes),  separated in energy-momentum space by a four-vector $b_{\mu}$~\cite{Armitage2017}.
A space- and time-dependent $b_{\mu}$, as in strained or inhomogenously magnetized Weyl semimetals~\cite{Cortijo2015} or Helium-3~\cite{Bevan1997,KV05}, generates chiral pseudomagnetic ($\mathbf{B}_5 = \nabla \times \mathbf{b}$) and pseudoelectric ($\mathbf{E}_5 = -\partial_t \mathbf{b} - \nabla b_0$) fields, which couple with opposite signs to opposite chiralities~\cite{Liu:2013kv,Chernodub2014,Schuster2016,PCF16,GVV16,Huang2017}.
These pseudo\-fields enhance or generalize phenomena ranging from transport to interface physics~\cite{PCF16,GVV16,Zhou2013,Sumiyoshi2016,GVV16,Cortijo2016b,Weststrom2017,Gorbar2017e,Gorbar2017f,GMS17b,Roy2017,
Yang2015,Jiang2015,Cortijo2016}.
Unlike $\mathbf{B}$, which generates Landau levels dispersing in opposite directions for opposite chiralities, $\mathbf{B}_5$ generates pseudo-Landau levels that disperse in the \textit{same} direction for both chiralities~\cite{Land16,Liu:2013kv,Chernodub2014,PCF16,GVV16}.

Applying the Landau level picture to chiral fields leads to puzzling conclusions:
for example, because of the chirality independent dispersion of the zeroth pseudo-Landau level due to $\mathbf{B}_5$, $\mathbf{E}$ increases (or decreases) the number of fermions for \textit{both} chiralities at a rate proportional to $\mathbf{E}\cdot\mathbf{B}_5$, giving an apparent nonconservation of the total charge.
This is expressed as the so-called \textit{covariant} anomaly~\cite{Land16,bertlmann2000anomalies}
\begin{align}
\label{eq:conschiralcov}
\partial_{\mu}J^{\mu}_{5,\mrm{cov}} =& \frac{1}{2 \pi^2} \left(\mathbf{E} \cdot \mathbf{B}+\mathbf{E}_5 \cdot \mathbf{B}_5\right),\\
\label{eq:covan}
\partial_{\mu}J^{\mu}_{\mrm{cov}} =& \frac{1}{2\pi^2} \left(\mathbf{E}_5 \cdot \mathbf{B}+\mathbf{B}_5 \cdot \mathbf{E}\right).
\end{align}
Neither the covariant chiral ($J_{5,\mathrm{cov}}^{\mu} = J^{\mu}_{L,\mathrm{cov}} - J^{\mu}_{R,\mathrm{cov}}$) nor vector ($J^{\mu}_\mathrm{cov} = J^{\mu}_{L,\mathrm{cov}} + J^{\mu}_{R,\mathrm{cov}}$) currents are conserved.
In field theory, to explicitly restore charge conservation, the covariant currents are supplemented by Bardeen polynomials $\delta j^\mu$, which act like boundary conditions for the accumulated charge at the cut-off energy~\cite{L14,Gorbar2017a,B69,Gorbar2017b}.
This procedure defines the \textit{consistent} anomaly~\cite{Land16,bertlmann2000anomalies}
\begin{align}
\label{eq:conschiralcons} 
\partial_{\mu}J^{\mu}_{5} =& \frac{1}{2 \pi^2} \left(\mathbf{E} \cdot \mathbf{B}+\dfrac{1}{3}\mathbf{E}_5 \cdot \mathbf{B}_5\right), \\
\label{eq:consan}
\partial_{\mu}J^{\mu} =& 0.
\end{align}
The consistent anomaly conserves charge, and thus determines observables~\cite{Gorbar2017a,Gorbar2017c} by discarding unphysical responses~\cite{L14,Vazifeh:2013fe,Land16,Gorbar2017a}.
The Bardeen polynomials, however, sacrifice intuition of the covariant picture based on Landau levels and obfuscate the restoration of charge conservation in specific lattice implementations.

Despite the field theory of the consistent and covariant anomalies being well understood for a long time~\cite{Land16}, a simple physical picture of their origin on a lattice, with guaranteed charge conservation, is still missing.
In this work we provide such a picture using as building blocks the Landau and pseudo-Landau levels.
It leads to our two main results: first, we identify the Fermi arcs as a source for the covariant anomaly terms of Eq.~\eqref{eq:covan} and relate them to the Bardeen polynomials.
Second, we show that when $\mathbf{B}_5>\mathbf{B}$ the Fermi surface twists into a bowtie shape, a property central to our understanding of how the term $\mathbf{E}\cdot\mathbf{B}_5$ redistributes charge within the sample. 
Our picture allows us to address as well the 1/3 disparity between the second term in Eq.~\eqref{eq:conschiralcons} and Eq.~\eqref{eq:conschiralcov}.
We argue that a necessary condition for its identification is a $\mathbf{B}_5$ profile that spatially separates chiral charge creation and annihilation.
Similar to how Landau levels simplified our understanding of the chiral anomaly~\cite{NielNino83}, we use pseudo-Landau levels (developed in Ref.~\onlinecite{GVV16}) to provide a unified and simple lattice picture of the consistent and covariant anomalies with specific implications for experiment.

%
Our starting point is the Weyl semimetal model~\cite{Vazifeh:2013fe}
\begin{align}
 \mathcal{H} =& v \left[ \sin( k_y) \sigma_x - \sin( k_x) \sigma_y \right] \tau_z + v\sin(  k_z) \tau_y + m \tau_x  \nonumber \\
 & + t \sum_i \left[ 1 - \cos( k_i )\right] \tau_x + v \sum_\mu u^\mu b_\mu, 
 \label{eq:weyl_lattice}
\end{align}
with $a=1$ the lattice constant, $\sigma_i$ ($\tau_i$) spin (orbital) Pauli matrices,
and $u^\mu =( \sigma_z \tau_y, -\sigma_x \tau_x, -\sigma_y \tau_x , \sigma_z )$.
For small $b_\mu=(b_0,\mathbf{b})$ and $m^2<v^2 | \mathbf{b}^2 - b_0^2|$, the model has one pair of Weyl nodes near $\Gamma$~\cite{Vazifeh:2013fe,GVB15}. 
Unless stated otherwise, we set $m=0$ and $t=2 v/\sqrt{3}$.
When $\mathbf{b}$ is oriented along a reciprocal lattice vector, this parameter choice gives two Weyl nodes located at $\pm \mathbf{b} [1+\mathcal{O}(b_0^2)] + \mathcal{O} (b_j^5) $ and energies $\pm v b_0 [1 + \mathcal{O} (b_j^2)] + \mathcal{O} (b_0^3)$.
Our results also apply to generalizations of Eq.~\eqref{eq:weyl_lattice} that model Dirac (e.g., Cd$_3$As$_2$, Na$_3$Bi) and Weyl (e.g., TaAs family) materials~\cite{Wang2012,Wang:2013is,Weng:2015ec}.
We further define $\gamma^\mu = (\tau_x, i \sigma_y \tau_y , -i \sigma_x \tau_y, i \tau_z)$ and $\gamma^5 = i \gamma^0 \gamma^1 \gamma^2 \gamma^3 = \sigma_z\tau_y$, such that $u^\mu = \gamma^0 \gamma^\mu \gamma^5$
and the space-dependent chiral charge density is
\begin{equation}
\label{eq:chcharge}
J_5^{0}(\mathbf{x}) \equiv \rho_5 (\mathbf{x}) = \sum_{n \in \mathrm{occ.}} \bra{\psi_n (\mathbf{x})} \gamma^5 \ket{\psi_n (\mathbf{x})}.
\end{equation}
The charge density, $\rho (\mathbf{x})$, is obtained by replacing $\gamma^5$ with the identity in Eq.~\eqref{eq:chcharge}.
While our lattice model~\eqref{eq:weyl_lattice} includes a $\tau_x$ term that explicitly breaks 
conservation of $\rho_5$, we show in the Supplemental Material how this effect is controlled.
\begin{figure}
\includegraphics[width=\columnwidth]{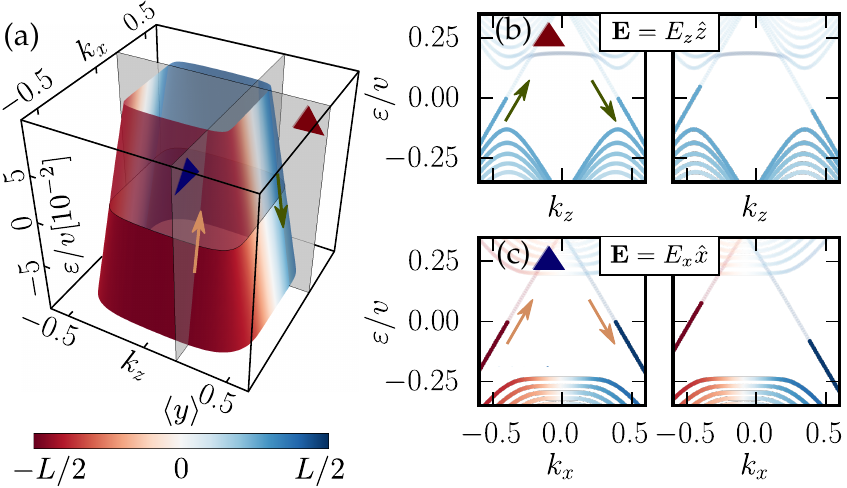}
\caption{Anomalies due to $\mathbf{E}\cdot\mathbf{B}$ and $\mathbf{E}\cdot\mathbf{B}_5$.
(a) The spectrum of Hamiltonian~\eqref{eq:weyl_lattice} for a slab finite along $y$, showing the coexisting Fermi arcs and Landau levels 
of $\mathbf{B}=B \hat{z}$.
The color scale denotes the wave function's position along $y$.
(b), (c) The Landau (pseudo-Landau) levels of $\mathbf{B}$ ($\mathbf{B}_5$)  on the red (blue) triangle plane show the anomaly in the presence of $\mathbf{E}=E_z \hat{z}$ ($\mathbf{E}=E_x \hat{x}$), where the spectral flow at the Fermi surface is shown by green (orange) arrows.
We implement $\mathbf{E}$ using the gauge choice $\mathbf{k} \to \mathbf{k} - \mathbf{E} t$.
The left panels show the occupation at $t=0$, while the right panels show the dispersion for $t>0$ with $E_{x,z} t = 0.08$ and the same momenta occupied as in the left panels, cf.\ Supplemental Material. 
Here $L=100$ and $\ell_B = 1/\sqrt{B} =11.2$.
}
\label{fig:unify}
\end{figure}

The Hamiltonian~\eqref{eq:weyl_lattice} (with $v=1$) derives from an effective field theory with action
\begin{equation}
 \mathcal{S} = \int d^4 x\, \bar{\psi}\left[\gamma^\mu \left(i\partial_\mu -A_\mu - b_\mu \gamma^5\right) - m \right] \psi,
 \label{eq:action}
\end{equation}
where $\bar{\psi} = \psi^\dagger \gamma^0$ and repeated indices are summed.
$\mathcal{S}$ yields two species of Weyl fermions of opposite chiralities, coupled to an external chiral field $b_\mu$ and a vector field $A_\mu$, and separated by $b_\mu$ for $m=0$.
The lattice regularization is given by the Wilson map $k_i \to \sin k_i$~\footnote{We restrict our analysis to isotropic Weyl nodes for simplicity.}, and $m \to m + t \sum_i \left( 1 - \cos k_i \right)$~\cite{Rothe}.

Spatial and temporal variations of $\mathbf{b}$ generate the chiral fields $\mathbf{B}_5=\nabla\times\bf{b}$ and $\mathbf{E}_5=-\partial_0\mathbf{b}$.
The simplest realization of $\mathbf{B}_5$ occurs at the boundary of any Weyl semimetal with vacuum, where the Weyl node separation $b_\mu$ goes to zero~\cite{Chernodub2014,Land16,GVV16}.
For example, for a slab along $y$, $\mathbf{b}(y) = b_z [\Theta(y-L/2)-\Theta(y+L/2)]\hat{z}$ gives $\mathbf{B}_5(y) = b_z[\delta(y-L/2)-\delta(y+L/2)]\hat{x},$ localized at the surface, which generates surface pseudo-Landau levels dispersing along $\pm k_x$, with opposite signs at each surface~\cite{GVV16}.
Their Fermi surface traces an arc, establishing the correspondence between surface pseudo-Landau levels induced by $\mathbf{B}_5$ and topological surface states. 
Analogously, a uniform external magnetic field $\mathbf{B}=B \hat{z}$ parallel to the Weyl node separation leads to a spectrum hosting bulk Landau levels dispersing along $\pm k_z$, where the sign is set by the Weyl node chirality. 
When both $\mathbf{B}$ and a surface $\mathbf{B}_5$ are present, Landau and pseudo-Landau levels coexist and the Fermi surface at the Fermi energy $\varepsilon_\mathrm{F}$ (set to $\varepsilon_\mathrm{F}=0$ hereafter) traces a rectangle~\footnote{The Fermi surface rectangle becomes a parallelogram for general orientations of $\bf{B}$}, see Fig.~\ref{fig:unify}(a)~\cite{BQ16,OK16}.
The coexistence of Landau and pseudo-Landau levels provides an ideal platform to discuss the anomalies.
Applying $\mathbf{E}=E_z \hat{z}$ pumps charges of one chirality to the other, connecting Landau levels of $\mathbf{B}$ through the band bottom and realizing the $\mathbf{E}\cdot\mathbf{B}$ term in Eq.~\eqref{eq:conschiralcons}~\cite{NielNino83} [see Fig.~\ref{fig:unify}(b)].
Similarly, since the pseudo-Landau levels disperse along $\pm k_x$ on each surface, applying $\mathbf{E}=E_x \hat{x}$ depletes charges from one surface and generate charges on the other [Fig.~\ref{fig:unify}(c)].
We can interpret~\cite{NielNino83} this as an anomaly of \textit{each} surface state due to $\mathbf{E}\cdot\mathbf{B}_5$.
In contrast to the chiral anomaly in the absence of chiral fields, where the total charge is locally conserved, the spatial separation of the two surfaces leads to an apparent violation of local charge conservation, as in Eq.~\eqref{eq:covan}.
Our pseudo-Landau level picture demonstrates the surface origin of the covariant anomaly, $\sim \mathbf{E}\cdot\mathbf{B}_5$.

Our picture tracks how charge is explicitly conserved.
The spectral flow between the pseudo-Landau levels at each surface happens via the bulk Landau levels connecting them, fixing local charge conservation.
The effect of $\mathbf{E}$ results in an adiabatic shift of $k_x$, which via position-momentum locking~\cite{GVV16} generates a Hall current
\begin{equation}
\label{eq:Hall}
\delta\mathbf{j} = -\frac{1}{2\pi^2} b_z E_x\hat{y}.
\end{equation}
We can interpret this as the net current flowing along $\hat{y}$ between surfaces through the bottom of the band. 
In the bulk, $b_z$ is constant leading to $\bm{\nabla}\cdot\delta\mathbf{j}=0$ and no accumulation of charge.
At the surface, the Weyl node separation varies, leading to a finite divergence of the spatial current, positive on one surface and negative on the other. 
\begin{figure}
\includegraphics[width=\columnwidth]{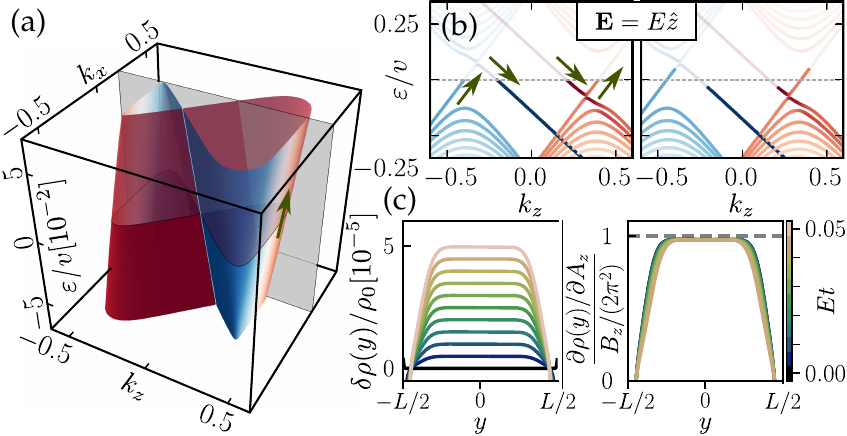}
\caption{
Anomaly due to $\mathbf{E}\cdot\mathbf{B}_5$ with a constant bulk $\mathbf{B}_5 = B_5 \hat{z}$.
(a) Energy spectrum with the same color coding and boundary conditions as Fig.~\ref{fig:unify}.
(b) The pseudo-Landau levels of $\mathbf{B}_5$ show an anomaly when $\mathbf{E}=E \hat{z}$ with the spectral flow indicated by the green arrows.
The left panel shows the occupation at $k_x = 0.2$ [gray plane in (a)] at $t=0$; in the right panel $E t = 0.05$.
(c) Left: The spatial profile of the charge $\delta \rho$ relative to that at $t=0$, normalized by the total charge $\rho_0$ for different times (colors), showing a surface to bulk charge redistribution.
The $y$ axis is chosen such that only the (positive) bulk contribution is visible, which is exactly canceled by the (negative) surface contribution.
Right: Derivative of the density with respect to the vector potential $A$ for different times, which equals $\partial_{A_z} \rho = B_5/(2\pi^2)$ in the bulk (dashed line).
The derivatives with respect to vector potential and time are related via $\partial_t \rho = -\mathbf{E} \cdot \partial_\mathbf{A} \rho$.
In (a) and (b), $L=100$ and $\ell_5 = 1/\sqrt{B_5} = 11.1$; in (c), $L=200$ and $\ell_5 = 15.8$.
}
\label{fig:edotb5}
\end{figure}

Similarly, more general profiles of  $\mathbf{B}$ and  $\mathbf{B}_5$ can be understood in terms of Landau and pseudo-Landau levels.
For instance, a uniform bulk $\mathbf{B}_5 \parallel \hat z$ arises from $\mathbf{b} = B_5 y\hat{x}$. 
Its spectrum [Fig.~\ref{fig:edotb5}(a)] shows a characteristic butterfly Fermi surface, obtained from the rectangular Fermi surface Fig.~\ref{fig:unify}(a) by noticing that the two bulk Landau levels have lengths $\mathbf{B}\pm\mathbf{B}_5$.
When $\mathbf{B}_5>\mathbf{B}$, the Fermi surface twists, leading to Fig.~\ref{fig:edotb5}(a) 
(found to lead to peculiar quantum oscillation signals in an unrelated context~\cite{BB17}).
An $\mathbf{E} = E\hat{z}$ parallel to $\mathbf{B}_5$ [Figs.~\ref{fig:edotb5}(b) and (c)] makes the bulk gain charge above the Fermi level (upward arrows), while the surface loses charge (downward arrows). 
This is consistent with the lattice numerics [Fig.~\ref{fig:edotb5}(e)] where the spatial profile of the charge relative to that at $t=0$ is shown for different times. 
\begin{figure}
\includegraphics[width=\columnwidth]{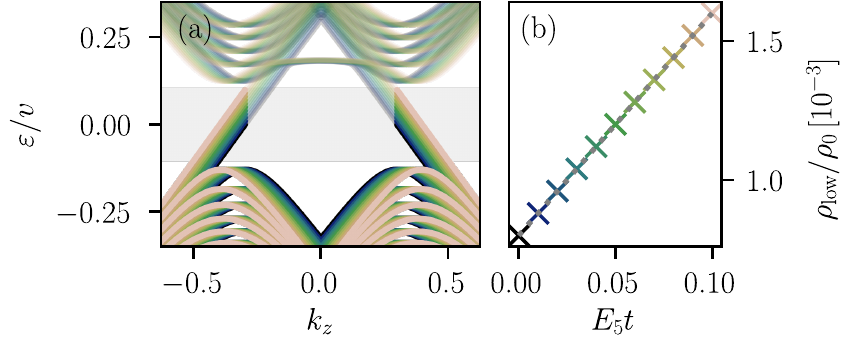}
\caption{
Anomaly due to $\mathbf{E}_5\cdot\mathbf{B}$. (a) Occupation at $k_x = 0.2$ and $\mathbf{B} \parallel \mathbf{E}_5 \parallel \hat{z}$ for different times [different colors, (b)], where dark (light) colors represent filled (empty) states and the gray shaded region is the low-energy regime, isolating the covariant anomaly.
We implement $\mathbf{E}_5$ by $\mathbf{b} \to \mathbf{b}- \mathbf{E}_5 t$ and use the same boundary conditions as Fig.~\ref{fig:unify} with $L=100$ lattice sites and $\ell_B = 11.2$.
(b) The charge in the gray low-energy regime $\rho_\mathrm{low}$ linearly increases with $E_5 t$.
Since the total charge is conserved, the charge in the band bottom decreases at the same rate (not shown).
}
\label{fig:e5dotb}
\end{figure}

Our previous examples (Figs.~\ref{fig:unify} and \ref{fig:edotb5}) are a consequence of the covariant anomaly which only considers the depletion and growth of charges at the Fermi level.
By construction, the corresponding current $J^{\mu}_\mathrm{cov}$ misses information from states away from the Fermi level,
and thus it is not conserved as dictated by Eq.~\eqref{eq:covan}, cf.\ Supplemental Material.
The consistent current, $J^{\mu}$, is obtained from the covariant current by adding the Bardeen polynomials $\delta j^\mu$
such that $\partial_{\mu}(J_\mathrm{cov}^{\mu}+\delta j^\mu)=\partial_{\mu}J^{\mu}=0$~\cite{L14,Gorbar2017a}.
Using Eq.~\eqref{eq:covan} and the definition of the pseudo-fields,
\begin{align}
\delta j^0 = \frac{1}{2\pi^2}\mathbf{b}\cdot \mathbf{B}, & &
\delta \mathbf{j} = \frac{1}{2 \pi^2}\left(b_0 \mathbf{B}-\mathbf{b}\times \mathbf{E}\right).
\label{eq:bardeenpoly}
\end{align}
Comparing \eqref{eq:bardeenpoly} to \eqref{eq:Hall} of our first example (Fig.~\ref{fig:unify}), we identify the latter as a part of the Bardeen polynomials~\cite{Land16}.
The benefit of the Landau level approach is its intuitive interpretation:
in the first example, $\mathbf{E}_5=0$ and the finite $\mathbf{E}\cdot\mathbf{B}_5$ pumps charge from one surface to another via the anomalous Hall effect [Eq.~\eqref{eq:Hall}] through $\partial_t J^0_\mrm{cov}= \mathbf{E}\cdot\mathbf{B}_5/(2\pi^2)$.
Our second example, Fig.~\ref{fig:edotb5}, can be interpreted similarly.
In the bulk charge grows as $\partial_t J^0_\mrm{cov}= \mathbf{E}\cdot\mathbf{B}_5/ (2 \pi^2)$ locally
while at the surface charge is depleted since $\mathbf{B}_5$ has the opposite sign.
The corresponding current~\eqref{eq:bardeenpoly} pumps charge from the surface to the bulk.
Since $\mathbf{b}=B_5 y \hat{x}$, there are local currents in the bulk $\delta \mathbf{j} \sim B_5 E y\hat{y}$ and $\bm{\nabla}\cdot\delta\mathbf{j} = \mathbf{B}_5\cdot \mathbf{E}/(2\pi^2)$ is precisely the growth rate of local charge [Fig.~\ref{fig:edotb5}(c)], reconciling the Fermi surface (covariant) picture with the charge  conserving (consistent) picture on the lattice~\footnote{Such bulk to surface charge transfer is a property of clean systems where the scattering time $\tau\to \infty$. For a finite amount of disorder, the charge buildup will be screened faster than it diffuses, but this regime is outside of the scope of this work.}. 

\begin{figure}
\includegraphics[width=\columnwidth]{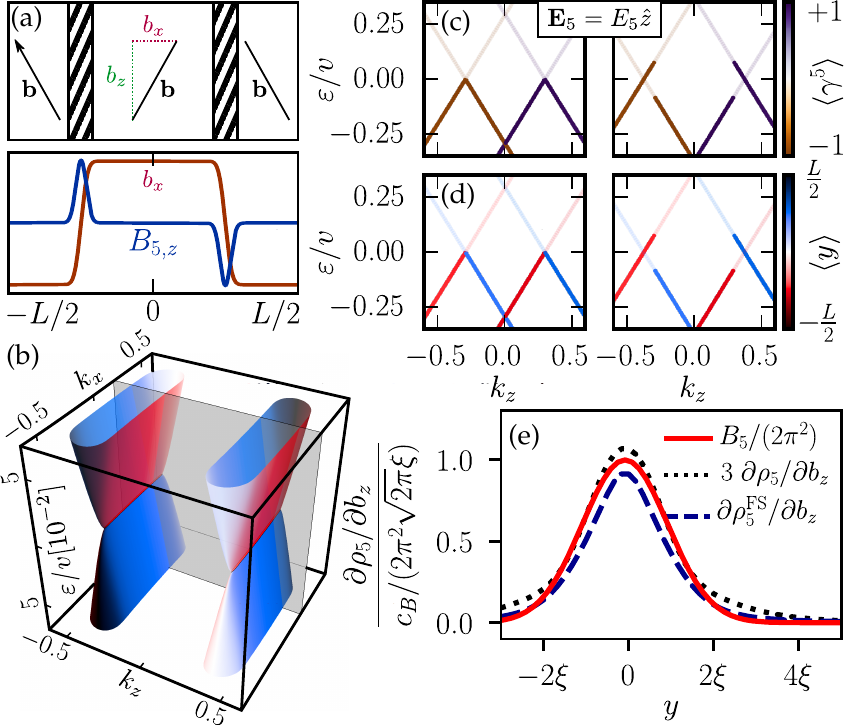}
\caption{
Anomaly due to $\mathbf{E}_5\cdot\mathbf{B}_5$.
(a) The Weyl node separation~\eqref{eq:bprof} and its corresponding $\mathbf{B}_5$.
(b) Energy spectrum with a real space color encoding that reflects the periodic boundary conditions.
(c), (d) Where $\mathbf{B}_5>0$ [red in (d)], a parallel $\mathbf{E}_5$ pushes left- and right-handed chiral charges above and below the Fermi level respectively, annihilating chiral charge.
Where $\mathbf{B}_5<0$ (blue), chiral charge is created.
(e) The spatial distribution of chiral charge creation and annihilation follows and equals $\mathbf{B}_5$ (red solid line) for charges $\rho_5^\mathrm{FS}$ traversing the original Fermi surface (blue dashed line).
With all bands, the total chiral charge creation is $\sim1/3$ of that at the Fermi surface, as predicted by the consistent anomaly (black dotted line).
We use $L = 360$, a $\mathbf{B}_5$ broadening $\xi=6$ and peak height $c_B = 1.25$, and a delta function broadening $\eta = 10^{-6}v $, cf.\ Supplemental Material.
}
\label{fig:e5dotb5}
\end{figure}

Moving on to spatio-temporally varying Weyl node separations, consider first $\mathbf{b} = -E_5 t\hat{z}$, yielding uniform $\mathbf{E}_5\parallel \hat{z}$.
A $\mathbf{B}\neq 0$ activates the second term of~\eqref{eq:covan}, suggesting that charge is created at the Fermi surface at a rate $\mathbf{E}_5\cdot\mathbf{B}/(2\pi^2)$. 
For our lattice model~\eqref{eq:weyl_lattice}, $\mathbf{E}_5$ shifts the band bottom, pushing charge above a fixed energy [Fig.~\ref{fig:e5dotb}(a)].
Rigid shifts of the band conserve total charge, giving the consistent picture of the anomaly.
However, if one insists on only considering the low-energy gray region in Fig.~\ref{fig:e5dotb}(a), the charge appears to emerge from the vacuum, as expected from the covariant anomaly.
The connection between these pictures is shown in Fig.~\ref{fig:e5dotb}(b), where charge growth near the Fermi surface equals charge loss near the band bottom, which in turn equals the Bardeen polynomial $\delta j^0$, Eq.~\eqref{eq:bardeenpoly}.

We end by addressing the factor of $1/3$ disparity between the prefactors of the $\mathbf{E}_5 \cdot \mathbf{B}_5$ anomalies.
This difference implies that the band bottom current must add $-2/3$ to the Fermi surface contribution, irrespective of the precise pseudo-field profile.
One may argue that this factor arises from the topological nature of the Bardeen polynomials~\cite{Gorbar2017b}, yet we find that topology alone does not explain the conditions which give $1/3$ for a generic lattice model.
To illustrate the conditions for isolating the $1/3$, consider~\cite{Roy2017}
\begin{equation}
\label{eq:bprof}
\mathbf{b} = \frac{c_B}{2} \left[ \mathrm{erf} \left( \frac{y-\frac{L}{4}}{\sqrt{2} \xi} \right)-\mathrm{erf} \left( \frac{y + \frac{L}{4}}{\sqrt{2} \xi} \right) -1 \right]\hat{x} + b_z\hat{z},
\end{equation}
with $c_B$ a constant and $\mathrm{erf} (x)$ the error function [Fig.~\ref{fig:e5dotb5}(a)].
In a periodic system of length $L$ in the $y$ direction, this profile realizes interfaces at $\pm L/4$
between two Weyl semimetals with different node separations, connected by regions where $\mathbf{b}$ smoothly changes to give
$\mathbf{B}_5 = B_5 (y) \hat{z}$ with $B_5 (y) = \frac{c_B}{\sqrt{2 \pi} \xi} [ e^{-(y-\frac{L}{4})^2/(2 \xi^2)} - e^{-(y+\frac{L}{4})^2/(2 \xi^2)} ]$.
This profile maintains the useful property that the two chiralities (eigenstates of $\gamma^5$) are well separated in momentum space [Fig.~\ref{fig:e5dotb5}(c)].
Adding $\mathbf{E}_5 \parallel \mathbf{B}_5$ to produce a chiral anomaly, chiral charge~\eqref{eq:chcharge} is created/annihilated in spatially separated regions [Fig.~\ref{fig:e5dotb5}(d)], at a rate that closely follows the spatial profile of $\mathbf{B}_5$ as expected from Eq.~\eqref{eq:conschiralcons} [Fig.~\ref{fig:e5dotb5}(e)].
The Fermi surface contribution to the chiral charge is calculated as
\begin{equation}
 \frac{d \rho_5^{FS} (y)}{d \mathbf{b}} = \sum_n \bra{\psi_n} \gamma^5 \Pi_y \ket{\psi_n} \bra{\psi_n} \partial_\mathbf{b} \mathcal{H} \ket{\psi_n} \delta( \mu - \varepsilon_n) .
\end{equation}
with $\Pi_y$ the projector to the position $y$, cf.\ Supplemental Material.
Comparing the chiral charge pumped at the Fermi surface and that in the full spectrum to the expected value proportional to $B_5$, we see that they are close, but neither perfectly reproduces the field theory prediction.

To determine the factors behind this mismatch we note that the erf profile minimizes the spatial overlap between regions of chiral charge creation and annihilation ($\partial\rho_5/\partial b_z >0$ and $<0$ at $y\approx+L/4$ and $y\approx-L/4$ respectively).
However, for both chiralities, there exist regions in momentum space [light/dark regions in Fig.~\ref{fig:e5dotb5}(b)] where the states are not well-localized in real space.
These regions spread out with increasing $\xi$, such that the erf profile evolves into a sinusoid-like function, resulting in a poorer match between the $B_5(y)$ profile and the chiral charge response, cf.\ Supplemental Material.
The sensitivity of the result to these regions, highlighted by the failure of the sinusoid-like profile where no clear spatial separation exists, implies that the value $1/3$ is modified by such effects for generic profiles of $\mathbf{B}_5$.
This detrimental overlap is minimized if $L \gg \xi \gg a$ is satisfied, allowing the intriguing possibility that an exact 1/3 may be recovered in this limit.
Finally, all anomaly terms present finite size and quadratic corrections to the low-energy field theory~\eqref{eq:action}, which we discuss in the Supplemental Material.

In summary, we have provided an intuitive lattice picture that is based on Landau and pseudo-Landau levels and connects the covariant and consistent anomalies.
We explicitly identified the Hall current as the Bardeen polynomial that connects the covariant anomalies of the Fermi arcs and restores charge conservation, most notably when the Fermi surface knots into a bowtie.
We expect that the bowtie Fermi surface and its response to external fields will be important to understand strained Weyl semimetals experimentally, in particular their transport properties.

Our work highlights that measuring the consistent or covariant anomaly  (e.g., the factor of 1/3) depends on whether the experimental probe is sensible to only the Fermi surface or rather the entire Fermi sea.
Additionally, perturbations such as strain allow other model parameters to depend on position, e.g, the Fermi velocity, as well as additional terms in Eq.~\eqref{eq:action}~\cite{Arjona2018}.
Our work motivates the study of these questions to interpret incipient experiments in strained Weyl semimetals.

\begin{acknowledgments}
\textit{Acknowledgements.-} We thank B.~Bradlyn, Y.~Ferreiros, R.~Ilan, F.~de~Juan, K.~Landsteiner and D.~Pesin for insightful discussions, and especially N.G.~Goldman for discussions that initiated this work and comments on the manuscript.
This research was funded by the Marie Curie Program under EC Grant Agreement No. 653846 (A.G.G.), the ERC Starting Grant No.~679722 and the Knut and Alice Wallenberg Foundation 2013-0093 (J.H.B.), the EPSRC Grant No. EP/N01930X/1 (S.R.), and supported by
the Laboratory Directed Research and Development (LDRD) funding from Lawrence Berkeley National Laboratory, provided by the Director, Office of Science, of the U.S. Department of Energy under Contract No. DEAC02-05CH11231, and the DOE Basic Energy Sciences (BES) TIMES initiative as well as UTD Research Enhancement Funds (M.H.K.). 
\end{acknowledgments}

\appendix

\section{Supplemental Material}

\section{Summary of consistent and covariant anomalies}

For completeness, we provide a summary of the relevant facts of consistent and covariant anomalies derived from quantum field theory.
As discussed in the main text electric and magnetic fields that satisfy $\mathbf{E}\cdot\mathbf{B}\neq 0$ generate an anomalous imbalance between left and right movers.
Additionally pseudo-magnetic ($\mathbf{B}_5$) and pseudo-electric ($\mathbf{E}_5$) field---the fields that couple with opposite signs to opposite chiralities---will result in the same chiral imbalance, since the contributions of the two chiralities to the chiral current $J_{5,\mathrm{cov}}^{\mu} = J^{\mu}_{L,\mathrm{cov}} - J^{\mu}_{R,\mathrm{cov}}$ ($\mu=0,1,2,3$) will add up.
However, these considerations imply also that the vector current $J^{\mu}_\mathrm{cov} = J^{\mu}_{L,\mathrm{cov}} + J^{\mu}_{R,\mathrm{cov}}$ is not conserved.
Mathematically
\begin{subequations}
\begin{align}
\label{eq:cov3app}
  \partial_{\mu}J^{\mu}_{5,\mathrm{cov}} &= \dfrac{1}{2\pi^2} \left(\mathbf{E} \cdot \mathbf{B}+\mathbf{E}_5 \cdot \mathbf{B}_5\right),\\
  \label{eq:cov4app}
  \partial_{\mu}J^{\mu}_\mathrm{cov} &= \dfrac{1}{2\pi^2} \left(\mathbf{E} \cdot \mathbf{B}_5+\mathbf{E}_5 \cdot \mathbf{B}\right),
\end{align}
\label{eq:cov_app}%
\end{subequations}
a result which can be obtained either diagrammatically~\cite{bertlmann2000anomalies,L14} or semiclassically~\cite{Liu:2013kv,PCF16,GVV16,Gorbar2017a}.
Eqs.~\eqref{eq:cov_app} define the \textit{covariant} chiral anomaly \cite{L14,Land16} of the covariant vector and axial currents $J^{\mu}_\mathrm{cov}$ and $J^{\mu}_{5,\mathrm{cov}}$ respectively.

This result is troublesome: a theory which does not conserve charge is unphysical, yet
the fields $\mathbf{B}_5$ and $\mathbf{E}_5$ can be physically realized in the solid state~\cite{Liu:2013kv,PCF16,GVV16}.
At the quantum field theory level curing the nonconservation of charge amounts to acknowledging that Eqs.~\eqref{eq:cov_app} are determined solely by the Fermi surface, but that counteracting currents are necessary at energies comparable to the cut-off energy~\cite{L14}.
To ensure this constraint is satisfied, two additional Chern-Simons currents, $\delta j^\mu$ and $\delta j^\mu_5$, known as Bardeen polynomials, are added, which transfer the necessary charge and guarantee charge conservation \cite{B69}.
They relate the covariant currents ($J^\mu_\mathrm{cov},J^\mu_{5,\mathrm{cov}}$) to the physical consistent currents (${J}^\mu,{J}^\mu_5$) through ${J}^\mu_5 = J^\mu_{5,\mathrm{cov}} + \delta j^\mu_5$ and $J^\mu = J^\mu_\mathrm{cov} + \delta j^\mu$, respectively, and take the form
\begin{align}
\label{eq:bardeenpolyapp}
\delta j^0 &= \frac{1}{2\pi^2} \mathbf{A}_{5}\cdot \mathbf{B}; & & \delta \mathbf{j} = \frac{1}{2 \pi^2}\left(A_{5}^{0} \mathbf{B}-\mathbf{A}_{5}\times \mathbf{E}\right) ; \\
\label{eq:bardeenpoly5app}
\delta j^0_5 &= \frac{1}{6\pi^2}\mathbf{A}_{5}\cdot \mathbf{B}_5; & & \delta \mathbf{j}_5 =  \frac{1}{6\pi^2}\left(A_{5}^{0} \mathbf{B}_5-\mathbf{A}_{5}\times \mathbf{E}_5\right) ,
\end{align}
where the chiral gauge field $A^{\mu}_5$ defines the pseudo-fields $\mathbf{B}_5=\diff \times \mathbf{A}_{5} $ and $\mathbf{E}_5= -\diff A^0_{5}-\partial_0 \mathbf{A}_{5}$.
Together with Eqs.~\eqref{eq:cov_app} results in the \textit{consistent} anomaly~\cite{L14,Gorbar2017a}
\begin{align}
\label{eq:const}
\partial_{\mu} J^{\mu}_5 = \dfrac{1}{2\pi^2} \left(\mathbf{E} \cdot \mathbf{B}+\dfrac{1}{3}\mathbf{E}_5 \cdot \mathbf{B}_5\right); & &
\partial_{\mu} J^{\mu}=0.
\end{align}
Eqs.~\eqref{eq:cov_app} and~\eqref{eq:const} summarize the two versions of the quantum anomalies.

We note that the consistent and covariant currents have been discussed in the context of the two-dimensional quantum Hall effect~\cite{Stone:2012kd}. In this case the edge current is the sum of the edge consistent current plus the anomaly inflow from the 2+1D bulk Chern Simons term, which combined form the covariant edge current~\cite{Chandra94}. In a similar way, consistent and covariant currents can be defined for a 3+1D Weyl fermion living at the edge of a 4+1D  quantum Hall effect.

\section{Details of the lattice calculations}
\label{sec:details_lattice}

We are interested in the response of a Weyl semimetal to external fields $\mathbf{E}$ and $\mathbf{E}_5$.
A time-dependent vector potential $\mathbf{A} = -\mathbf{E} t$ gives an electric field, and analogously a time-dependent node separation $\mathbf{b} = -\mathbf{E}_5 t$.
We express the time-derivative of any expectation value $f$ in terms of the response to changes in $\mathbf{A}$ and $\mathbf{b}$,
\begin{equation}
 \frac{d f}{d t} = -\mathbf{E} \cdot \partial_\mathbf{A} f - \mathbf{E}_5 \cdot \partial_\mathbf{b} f ,
\end{equation}
which is especially handy when evaluating time-derivatives proportional to $\mathbf{E}$ and $\mathbf{E}_5$, such as $\partial_\mu j^\mu$ and $\partial_\mu j_5^\mu$ where $f$ corresponds to the densities $\rho$ or $\rho_5$.

The response of the density to external fields $\mathbf{E}$ and $\mathbf{E}_5$ is the response to a shift $\mathbf{A} \to \mathbf{A} + d \mathbf{A}$ and $\mathbf{b} \to \mathbf{b}+ d\mathbf{b}$ respectively.
Explicitly, the chiral density at site $y$ reads
\begin{equation}
 \rho_5 (y) = \sum_{n \in \mathrm{occ.}} \bra{\psi_n} \gamma^5 \Pi_y \ket{\psi_n},
\end{equation}
where $\Pi_y$ projects onto site $y$.
When performing an infinitesimal shift of the Weyl node separation $\mathbf{b} \to \mathbf{b} + d\mathbf{b}$, the single-particle states change as
\begin{equation}
 \ket{\psi_n (\mathbf{b} + d\mathbf{b})} = \ket{\psi_n} + d \mathbf{b} \cdot \sum_{m \neq n} \frac{\ket{\psi_m} \bra{\psi_n} \partial_\mathbf{b} \mathcal{H} \ket{\psi_m}}{\varepsilon_n - \varepsilon_m}
\end{equation}
resulting in a change of the chiral density
\begin{equation}
 \frac{d \rho_5 (y)}{d \mathbf{b}} = \sum_{n \in \mathrm{occ.}} \sum_{m \neq n} \frac{\bra{\psi_n} \gamma^5 \Pi_y \ket{\psi_m} \bra{\psi_n} \partial_\mathbf{b} \mathcal{H} \ket{\psi_m}}{\varepsilon_n - \varepsilon_m} + \mathrm{h.c.}
 \label{eq:response_all_energies}
\end{equation}
Note that the sum goes over the states $n$ that are initially occupied at a node separation $\mathbf{b}$.
The Fermi surface contribution, on the other hand, just counts the subset of states that are lifted above the Fermi level when shifting $\mathbf{b} \to \mathbf{b} + d\mathbf{b}$, minus the states that are pushed below the Fermi level, $d \rho_5^{FS} = d \rho_5^+ - d \rho_5^-$ with
\begin{align}
 d \rho_5^+ (y) &= \sum_{\substack{ \varepsilon_n <\mu, \\ \varepsilon_n + d \varepsilon_n >\mu}} \bra{\psi_n} \gamma^5 \Pi_y \ket{\psi_n} \\
 d \rho_5^- (y) &= \sum_{\substack{ \varepsilon_n >\mu, \\ \varepsilon_n + d \varepsilon_n <\mu}} \bra{\psi_n} \gamma^5 \Pi_y \ket{\psi_n}
\end{align}
or, expressed in terms of Heaviside functions,
\begin{align}
 d \rho_5^\pm (y)
 =& \pm \sum_n \bra{\psi_n} \gamma^5 \Pi_y \ket{\psi_n} \Theta ( \pm d \varepsilon_n) \nonumber \\
 & \times \left[ \Theta (\mu - \varepsilon_n) - \Theta ( \mu - (\varepsilon_n + d \varepsilon_n)) \right] .
\end{align}
To first order in perturbation theory, the infinitesimal change in the energies is
\begin{equation}
 d \varepsilon_n = d \mathbf{b} \cdot \bra{\psi_n} \partial_\mathbf{b} \mathcal{H} \ket{\psi_n}
\end{equation}
and the sum of Heaviside functions reduces to
\begin{equation}
 \Theta (\mu - \varepsilon_n) - \Theta (\mu - (\varepsilon_n + d \varepsilon_n)) = d \varepsilon_n \delta (\mu - \varepsilon_n),
\end{equation}
which finally gives
\begin{equation}
 \frac{d \rho_5^{FS} (y)}{d \mathbf{b}} = \sum_n \bra{\psi_n} \gamma^5 \Pi_y \ket{\psi_n} \bra{\psi_n} \partial_\mathbf{b} \mathcal{H} \ket{\psi_n} \delta( \mu - \varepsilon_n)
 \label{eq:response_fermi_surface}
\end{equation}
as used in the main text.
Numerically, we use a Lorentz approximation to the delta function, $\delta (x) = \lim\limits_{\eta \to 0} \eta/(\pi (\eta^2 + x^2))$.

\section{The quest for $1/3$}
\label{sec:appendix_quest}

\begin{figure}
\includegraphics{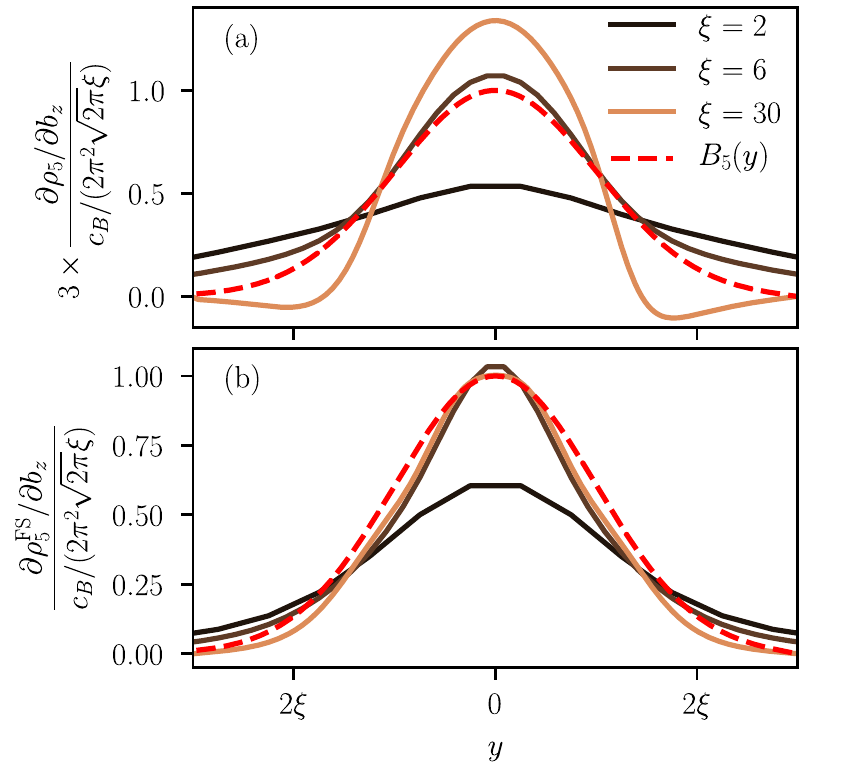}
\caption{(a) Responses $\partial \rho_5 /\partial b_z$ and (b) $\partial \rho_5^\mathrm{FS} /\partial b_z$ for our lattice model [Eq.~\eqref{eq:weyl_lattice}] with periodic boundary conditions in $y$-directions of length $L=360$ and $c_B=1.25$, cf.\ Eq.~\eqref{eq:bprof}.
  We show results for different widths $\xi$ of the regions with non-zero $\mathbf{B}_5$.
  The best matching of the field theory predictions are found for $\xi=6$, which mostly satisfies the condition $a \ll \xi \ll L$.
  For $\xi=2$, the condition $a \ll \xi$ is not satisfied, making lattice scale effects extremely relevant and deviating from the field theory limit.
  For $\xi=30$, tails of the two chiral charge pumping regions centered at $\pm L/4$ overlap, leading to notable finite size effects.
}
\label{fig:appendix_vary_xi}
\end{figure}

Two effects challenge the calculation of the chiral anomaly due to $\mathbf{E}_5 \cdot \mathbf{B}_5$ on a lattice.
First, $\mathbf{B}_5$ always averages to zero over the whole sample~\cite{GVV16}, which implies that a length scale $\xi$ exists that characterizes the typical size of the variations in $\mathbf{B}_5$.
Second, a mass term $m$ changes the response of the chiral charge $\partial_\mu j_5^\mu$.

\subsubsection{Effect of $\xi$}

We demonstrate the lattice effect set by the length $\xi$ explicitly by using the same profile of the node separation $\mathbf{b}$ as in the main text~\eqref{eq:bprof}.
In Fig.~\ref{fig:appendix_vary_xi} (a) and (b), we show how the responses of the total chiral charge $\rho_5$ and the charge around the Fermi surface $\rho_5^\mathrm{FS}$ are affected by different sizes of $\xi$ respectively.
The two conditions for $\xi$ discussed in the main text that allow for the identification of $1/3$ are both demonstrated explicitly.
When $\xi \approx a$ the variation of $B_5$ occurs on the scale of the lattice constant forcing the density variations Eqs.~\eqref{eq:response_all_energies} and \eqref{eq:response_fermi_surface} to interpolate smoothly between lattice sites such that the quantum field theory predictions breaks down.
When $\xi \ll L$ is not satisfied, tails of the regions with positive and negative $B_5(y)$ (positive and negative chiral charge creation) overlap, leading to notable finite size effect and again conflicting with the field theory predictions.

\begin{figure}
\includegraphics{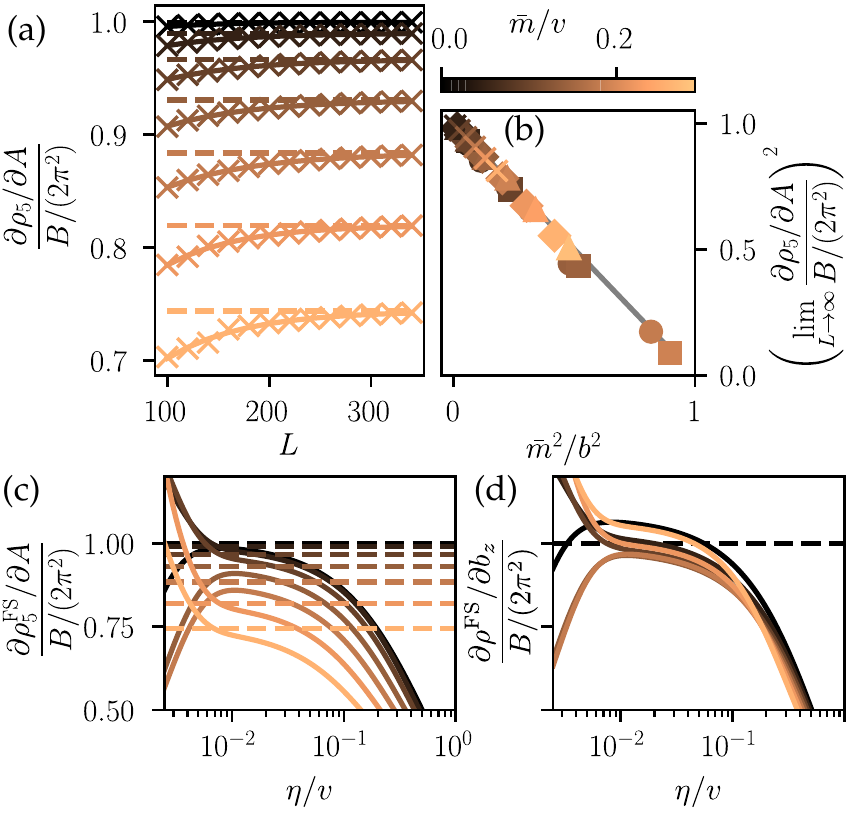}
\caption{(a) Convergence of the response $\partial \rho_5 /\partial A_z$ for a tight-binding system governed by the Hamiltonian~\eqref{eq:weyl_lattice} with periodic boundary conditions in $y$-directions of length $L$ and a node separation $\mathbf{b}=b \hat{z}$.
A magnetic field $\mathbf{B} = B \hat{z}$ that satisfies periodic boundary conditions is included, with a magnetic length $\ell_B = \sqrt{ 2\pi L}$.
The crosses denote the numerical results from tight-binding calculations , the solid line the results of a fit with $f(L) = c_0 + c_1 \exp (-L/\xi)$, and the dashed line the coefficient $c_0$, i.e., the result in the limit $L \to \infty$.
The color denotes different values of $\bar{m}$, as shown in panel~(b), and we further choose $b=0.4$.
(b) The response $\partial \rho_5 /\partial A_z$ in the limit $L \to \infty$ as a function of $\bar{m}^2/b^2$.
The different symbols denote different combinations of $b$ and $t$ and the gray line shows the expectation from the lattice model~\eqref{eq:response_mbar}.
(c) Response of the chiral charge around the Fermi surface to $\mathbf{E}$ $\partial \rho^\mathrm{FS}_5 /\partial A$, as a function of the level broadening $\eta$ for $b=0.4$.
The dashed lines denote the result from panel~(a), $\lim_{L \to \infty} \partial \rho_5/\partial A$.
(d) Response of the total charge around the Fermi surface to an axial field, $\partial \rho^\mathrm{FS} /\partial A_5$, as a function of the level broadening $\eta$.
The dashed line denotes the field theory result $\partial \rho/\partial A_5 = B/(2\pi^2)$ as a guide for the eyes.
}
\label{fig:convergence_EB}
\end{figure}

\begin{figure}
\includegraphics{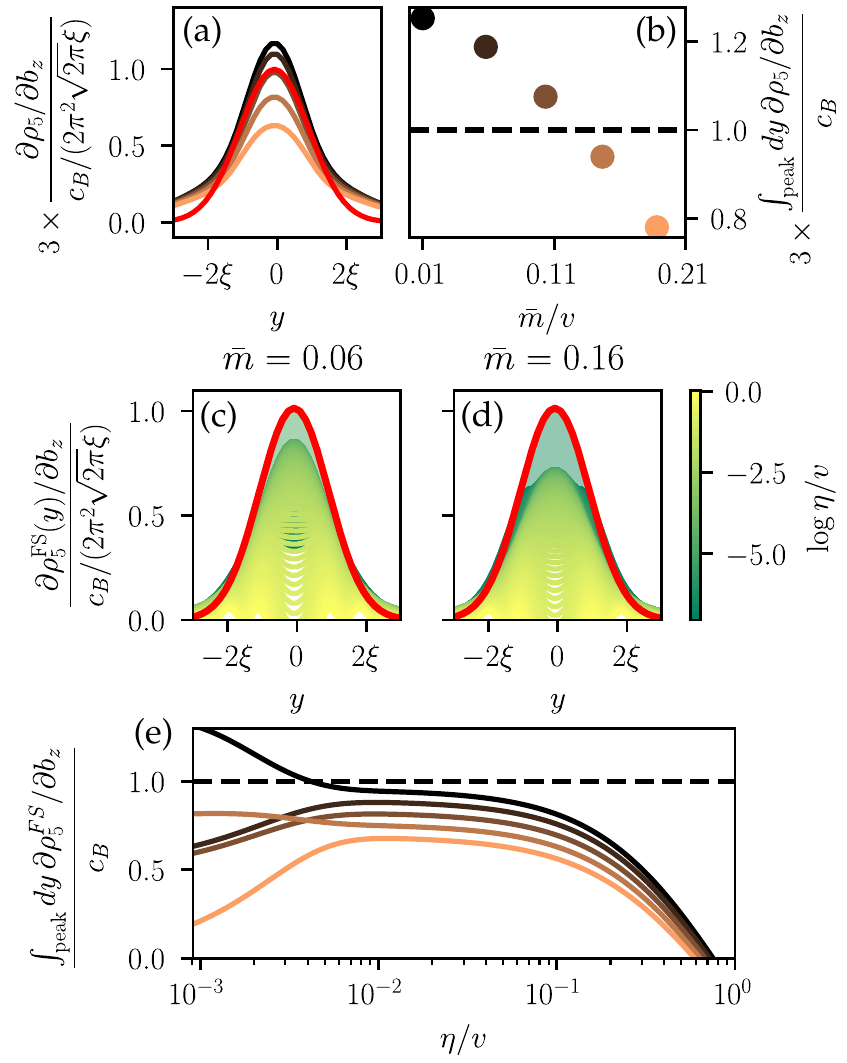}
\caption{(a) Response $\partial \rho_5 /\partial b_z$ for a tight-binding system governed by the Hamiltonian~\eqref{eq:weyl_lattice} with periodic boundary conditions in $y$-directions of length $L=360$.
An axial field $\mathbf{B}_5 = B_5 (y) \hat{z}$ generated by $\mathbf{b} (y)$, Eq.~\eqref{eq:bprof}, with $\xi=6$ and $c_B = 1$ that satisfies periodic boundary conditions is included.
The different colors denote different values of $\bar{m}$ (cf.\ panel (b)) and the bold red line shows the profile of $B_5 (y)$.
(b) The integral over $y$ around the peak is computed numerically for different values of $\bar{m}$.
The dashed line denotes the value of the integral that is expected from quantum field theory.
In (c) and (d), we show the response of the chiral charge around the Fermi surface $\partial \rho^\mathrm{FS}_5 /\partial b_z$ as a function of the level broadening $\eta$ with the different colors denoting different values of $\eta$.
The bold red line shows the profile of $B_5 (y)$, the filled region is minimal difference between $B_5$ and $\partial \rho^\mathrm{FS}_5 /\partial b_z$.
In (e), we show the numerically computed integral over $y$ around the peak in $B_5$ for different values of $\bar{m}$.
The dashed line denotes the value of the integral that is expected from quantum field theory.
}
\label{fig:appendix_response}
\end{figure}

\subsubsection{Effect of a finite mass}

The response of the chiral and total charge to fields $\mathbf{E}$ and $\mathbf{E}_5$ computed on a lattice can deviate from the expectation based on the simplest quantum field theory due to corrections set by the mass term, as mentioned in the main text.
In this section, we carefully investigate one example, the electromagnetic contribution to the chiral anomaly on a lattice, $\partial_\mu j_5^\mu = \mathbf{E} \cdot \mathbf{B}/(2\pi^2)$ and argue how a mass term influences the chiral anomaly.
We further show that the mass term does not play the same role for the covariant anomaly $\partial_\mu {J}^\mu_{\mrm{cov}} = \mathbf{E}_5 \cdot \mathbf{B}/(2\pi^2)$. 

The main source of lattice corrections to the anomaly is that the Wilson fermion Hamiltonian~\cite{Wilson:1974ji} used for all tight-binding calculations, Eq.~\eqref{eq:weyl_lattice}, has a momentum-dependent mass term
\begin{equation}
 M_\mathbf{k} = m + t \sum_i (1 - \cos k_i),
\end{equation}
that ensures the absence of copies of the Weyl nodes, or doublers, on the lattice~\cite{Wilson:1974ji}.
When $t>0$, the minimum of $M_\mathbf{k}$ is at the $\Gamma$ point and its maximum  at $(\pi,\pi,\pi)$;
for $m=0$, as set in the main text, the term is zero at $\mathbf{k}=0$ and increases away from $\Gamma$.
The term $M_\mathbf{k}$ sets the momentum-independent mass term $m$ in the action~\eqref{eq:action} that couples both chiralities and changes the response to electric and magnetic fields~\cite{Jackiw:1999el,Perez:1999fj,G12}.
Such a term results in an additional classical contribution to the chiral anomaly, $\partial_\mu j_{5,\mathrm{class}}^\mu = 2 m i \bar{\psi} \gamma^5 \psi$~\cite{bertlmann2000anomalies}, that is zero in all equilibrium situations we consider, which we confirm numerically.
Instead, another consequence of $m$ affects the chiral anomaly:
when $m \neq 0$ in the field theory, the eigenfunctions of the corresponding Hamiltonian are no longer eigenfunctions of the chiral matrix $\gamma^5$.
Instead, close to the Weyl node of chirality $\chi$, the overlap with $\gamma^5$ for the low-energy eigenstates with a linear dispersion can be obtained analytically to be
\begin{equation}
 \bra{ \psi^\chi} \gamma^5 \ket{\psi^\chi} = \chi \sqrt{ 1- \frac{m^2}{|b^2|}},
\end{equation}
with $b^2 = b_\mu b^\mu$, i.e., the chiral density around the Weyl is reduced by a nonzero mass term.
We note that a similar factor appears in the effective action for this model and is related to the Weyl node separation~\cite{Perez:1999fj,G12}.

On a lattice, the eigenstates of the Hamiltonian~\eqref{eq:weyl_lattice} are only eigenstates of $\gamma^5$ for those momenta $\mathbf{k}$ where $M_\mathbf{k} = 0$.
In particular, the overlap with $\gamma^5$ of states of chirality $\chi=\pm$ close to the Weyl nodes at $\chi \mathbf{k}_W$ is
\begin{equation}
  \bra{ \psi^\chi_{\chi \mathbf{k}_W}} \gamma^5 \ket{\psi^\chi_{\chi \mathbf{k}_W}} = \chi \sqrt{ 1- \frac{M_{\mathbf{k}_W}^2}{|b^2|}},
\end{equation}
i.e., it scales with the value of $M_\mathbf{k}$ at the Weyl nodes, $\bar{m} \equiv M_{\mathbf{k}_W}$.
In the presence of $\bar{m}\neq0$ the eigenstates are not strictly eigenstates of $\gamma^5$ yet in order to define the chiral anomaly using Eq.~\eqref{eq:chcharge} we effectively assume that they are.
An alternative definition of chirality that does not possess this problem is to partition the Brillouin zone into two and declare left and right chiral charges as done in Ref.~\onlinecite{Roy2016}.
While it has the advantage of having a clear definition of chirality, it does not reduce to any representative matrix at the field theory level and so we retain the first definition.

With these definitions it is tempting to upgrade the chiral anomaly to
\begin{equation}
\partial_\mu j_5^\mu = \frac{1}{2\pi^2} \sqrt{1 - \frac{\bar{m}^2}{b^2}} \mathbf{E} \cdot \mathbf{B}.
\label{eq:response_mbar}
\end{equation}
It is interesting to note that a similar equation would be obtained if one interprets the prefactor as an effective renormalization of the chiral electric charge $e_5 = e\sqrt{1 - \bar{m}^2/b^2}$ 
that would enter the chiral anomaly throughout the coupling to an external chiral gauge field $e_5A_{\mu,5}$ in analogy with ordinary charge~\footnote{We thank F. de Juan for this observation.}.
We have unsuccessfully attempted to derive Eq.~\eqref{eq:response_mbar} from the effective field theory responses for the action Eq.~\eqref{eq:action} discussed in~\cite{Jackiw:1999el,Perez:1999fj,G12}.
However, our numerical results on the corrections presented below suggest a richer structure when the mass is non-zero, which has motivated us to leave the precise connection between the quantum field theory discussed in these works and our results for a separate study.
At the end of the section we specify the conditions where our numerics coincide with the massless field theory expectation validating our results in the main text, up to a given order in momentum.

To gain insight on the above expectation numerically, we investigate how the chiral anomaly changes with $\bar{m}$.
Differentiating between the total charge and the charge around the Fermi surface, we compute the response to external fields $\mathbf{E}$ and $\mathbf{E}_5$ via Eqs.~\eqref{eq:response_all_energies} and~\eqref{eq:response_fermi_surface} on a lattice with periodic boundary conditions and a constant $\mathbf{B}$.

We show the response of the chiral charge $\rho_5$ in Fig.~\ref{fig:convergence_EB}(a) and~(b).
In panel (a), we show the convergence with system size, whereas in panel~(b), we show that, in the limit of $L\to \infty$, the response of the chiral charge scales with $\sqrt{1-\bar{m}^2/b^2}$ as dictated by  Eq.~\eqref{eq:response_mbar}.

Taking into account just states around the Fermi surface via Eq.~\eqref{eq:response_fermi_surface}, the chiral charge $\rho_5^\mathrm{FS}$ responds approximately in the same way as shown in Fig.~\ref{fig:convergence_EB}(b), cf.\ panel (c).
Thus, the consistent and covariant anomalies are the same for the term $\mathbf{E} \cdot \mathbf{B}$, even in the presence of a mass $m$.
In panel (d), we show the response of the total charge around the Fermi surface to a field $\mathbf{E}_5$:
due to the covariant anomaly, $\partial \rho^\mathrm{FS} /\partial A_5 \neq 0$.
In contrast to $\partial \rho^\mathrm{FS}_5/\partial A$, the contribution $\partial \rho^\mathrm{FS}/\partial A_5$ is not much affected by $\bar{m} \neq 0$.

The chiral-symmetry breaking term $M_\mathbf{k}$ has more intricate consequences for the anomaly in the chiral charge due to $\mathbf{E}_5 \cdot \mathbf{B}_5$.
We find numerically that the dependence of $\partial \rho_5/\partial b_z$ and $\partial \rho^\mathrm{FS}_5/\partial b_z$ on $\bar{m}$ are different.
In Fig.~\ref{fig:appendix_response}(a), we show the response of the total chiral charge for different values of $\bar{m}$ with the profile of $\mathbf{b}$ according to Eq.~\eqref{eq:bprof}.
It allows us to obtain panel~(b), where the value of the integral over the curves in Fig.~\ref{fig:appendix_response}(a) as a function of $\bar{m}$ is compared to the result expected from the massless quantum field theory, i.e.
the 1/3 prefactor difference between the second term of equations \eqref{eq:bardeenpolyapp} and \eqref{eq:const} shown as a dashed line.
$\partial \rho_5 /\partial b_z$ is closest to the massless field theory prediction for a value  $\bar{m}\sim0.1v$, close to $m=0$.
This contrasts with the $\mathbf{E}\cdot\mathbf{B}$ analyzed above where the massless field theory expectation was reached when $\bar{m}=0$.
Finally, the effect of $\bar{m}$ on $\partial \rho^\mathrm{FS}_5/\partial b_z$ is shown in Fig.~\ref{fig:appendix_response}(c) and~(d) for various values of the level broadening $\eta$.
The integral over one peak decreases with $\bar{m}$, as visible in Fig.~\ref{fig:appendix_response}(e), different from the total charge contribution.

In conclusion, the massless quantum field theory expectation is recovered numerically when $\bar{m}\sim m \sim 0$. 
This is satisfied when quadratic corrections are negligible and we choose $m\sim0$,
a regime which is largely satisfied by our parameter choices in the main text. 
The small yet rich deviations that we have quantified numerically (Figs.~\ref{fig:convergence_EB} and~\ref{fig:appendix_response}) indicate that further effort beyond the scope of this work is required to precisely connect with quantum field theory expectations with a finite mass.

\bibliography{weyl_cc.bib}

\end{document}